\documentclass[preprint,12pt]{elsarticle}

\let\today\relax
\makeatletter
\def\ps@pprintTitle{%
    \let\@oddhead\@empty
    \let\@evenhead\@empty
    \def\@oddfoot{\footnotesize\itshape
         {} \hfill\today}%
    \let\@evenfoot\@oddfoot
    }
\makeatother

\usepackage{amssymb}
\usepackage{amsmath}
\usepackage{algorithm}
\usepackage{algorithmic}
\usepackage{stfloats}
\usepackage{float} 
\usepackage{color}

\usepackage[backref]{hyperref}
\hypersetup{
hidelinks}

\usepackage{natbib}

\usepackage{listings, xcolor}
\definecolor{verylightgray}{rgb}{.97,.97,.97}
\def\showauthors@{T}
\lstdefinelanguage{Solidity}{
	keywords=[1]{anonymous, assembly, assert, balance, break, call, callcode, case, catch, class, constant, continue, constructor, contract, debugger, default, delegatecall, delete, do, else, emit, event, experimental, export, external, false, finally, for, function, gas, if, implements, import, in, indexed, instanceof, interface, internal, is, length, library, log0, log1, log2, log3, log4, memory, modifier, new, payable, pragma, private, protected, public, pure, push, require, return, returns, revert, selfdestruct, send, solidity, storage, struct, suicide, super, switch, then, this, throw, transfer, true, try, typeof, using, value, view, while, with, addmod, ecrecover, keccak256, mulmod, ripemd160, sha256, sha3}, % generic keywords including crypto operations
	keywordstyle=[1]\color{blue}\bfseries,
	keywords=[2]{address, bool, byte, bytes, bytes1, bytes2, bytes3, bytes4, bytes5, bytes6, bytes7, bytes8, bytes9, bytes10, bytes11, bytes12, bytes13, bytes14, bytes15, bytes16, bytes17, bytes18, bytes19, bytes20, bytes21, bytes22, bytes23, bytes24, bytes25, bytes26, bytes27, bytes28, bytes29, bytes30, bytes31, bytes32, enum, int, int8, int16, int24, int32, int40, int48, int56, int64, int72, int80, int88, int96, int104, int112, int120, int128, int136, int144, int152, int160, int168, int176, int184, int192, int200, int208, int216, int224, int232, int240, int248, int256, mapping, string, uint, uint8, uint16, uint24, uint32, uint40, uint48, uint56, uint64, uint72, uint80, uint88, uint96, uint104, uint112, uint120, uint128, uint136, uint144, uint152, uint160, uint168, uint176, uint184, uint192, uint200, uint208, uint216, uint224, uint232, uint240, uint248, uint256, var, void, ether, finney, szabo, wei, days, hours, minutes, seconds, weeks, years},	% types; money and time units
	keywordstyle=[2]\color{teal}\bfseries,
	keywords=[3]{block, blockhash, coinbase, difficulty, gaslimit, number, timestamp, msg, data, gas, sender, sig, value, now, tx, gasprice, origin},	% environment variables
	keywordstyle=[3]\color{violet}\bfseries,
	identifierstyle=\color{black},
	sensitive=true,
	comment=[l]{//},
	morecomment=[s]{/*}{*/},
	commentstyle=\color{gray}\ttfamily,
	stringstyle=\color{red}\ttfamily,
	morestring=[b]',
	morestring=[b]"
}

\floatstyle{ruled}
\newfloat{listing}{tb}{lst}{}
\floatname{listing}{Listing}

\begin{document}

\bibliographystyle{plain}

\begin{frontmatter}



\title{Smart Contracts in the Real World: A Statistical Exploration of External Data Dependencies}


\author[label1]{Xiaoqi Li} %% Author name
\affiliation[label1]{organization={Hainan University},%Department and Organization
            addressline={csxqli@ieee.org}, 
            city={Haikou},
            country={China}}

\author[label3]{Yishun Wang} %% Author name
% Author affiliation
% \affiliation[label2]{organization={Hainan University},%Department and Organization
%             addressline={yishunwang@hainanu.edu.cn}, 
%             city={HaiKou},
%             country={China}}

\author{Yuqing Zhang}
            
\author{Shipeng Ye} %% Author name
% \affiliation[label3]{organization={Hainan University},%Department and Organization
%             addressline={yeshipeng35@gmail.com}, 
%             city={HaiKou},
%             country={China}}

\author{Lei Xie} %% Author name
% \affiliation[label4]{organization={Hainan University},%Department and Organization
%             addressline={xielei1916@gmail.com}, 
%             city={HaiKou},
%             country={China}}

\author{Ju Xing} %% Author name
% \affiliation[label5]{organization={Hainan University},%Department and Organization
%             addressline={sunsplanter@gmail.com}, 
%             city={HaiKou},
%             country={China}}

%% Abstract
\begin{abstract}
Smart contracts with external data are crucial for functionality but pose security and reliability concerns. Statistical and quantitative studies on this interaction are scarce. To address this gap, we analyzed 10,500 smart contracts, retaining 9,356 valid ones after excluding outdated or erroneous ones. We employed code parsing to transform contract code into abstract syntax trees and identified keywords associated with external data dependencies. We conducted a quantitative analysis by comparing these keywords to a reference list. We manually classified the 9,356 valid smart contracts to ascertain their application domains and typical interaction methods with external data. Additionally, we created a database with this data to facilitate research on smart contract dependencies. Moreover, we reviewed over 3,600 security audit reports, manually identifying 249 (approximately 9\%) related to external data interactions and categorized their dependencies. We explored the correlation between smart contract complexity and external data dependency to provide insights for their design and auditing processes. These studies aim to enhance the security and reliability of smart contracts and offer practical guidance to developers and auditors.
\end{abstract}


% \begin{highlights}
% \end{highlights}

%% Keywords

\end{frontmatter}



\section{Introduction}
The inherent decentralization of blockchain facilitates the execution of smart contracts with trust, eliminating the need for reliance on any single entity \citep{zhang2022authros}. This pivotal characteristic has substantially driven the broad adoption of blockchain technology, notably within the Ethereum ecosystem. Consequently, this ecosystem has spurred the creation of a multitude of decentralized applications (DApps) \cite{popchev2024decentralized,kostamis2024data,saian2024prototype} to meet the escalating demand for their functionality \citep{li2020characterizing}. The significance of external data for smart contracts is self-evident, given their frequent reliance on real-world data to enhance functionality. Engagement with real-time data, including prices or weather conditions, is commonplace for executing specific logic within smart contracts.\par
In this study, we examine 9,356 successfully compiled real-world smart contracts from an initial pool of 10,500, collected from Etherscan via a web crawler. We have excluded contracts that are outdated or contain syntax errors. Specifically, upon reviewing 240 smart contract source codes that interact with external data, we have compiled a keyword list titled "oracle\_services," encompassing pertinent contract and function names. Furthermore, employing Abstract Syntax Trees (ASTs) \citep{huang2024revealing,nguyen2024development}, a tree-like data structure extensively utilized in compilers and interpreters to represent the abstract syntax structure of programming language code. We then compared the 9,356 contracts against the "oracle\_services" list, determining the number and proportion of contracts that involve external data interaction. \par
Additionally, we have selected 249 smart contract security audit reports from three security teams to evaluate their interactions with external data. The level of dependency on external data is classified as low, medium, or high, based on the frequency of interaction. Increased dependency on external data correlates with a higher risk of security vulnerabilities associated with such data.\par
The main contributions of this paper are as follows:

\begin{itemize}
    \item To the best of our knowledge, this paper presents the first empirical study on the dependence of smart contracts on external data. The aim is to assist in reducing security issues caused by smart contract developers when interacting with external data.
    \item We analyze real-world smart contracts, quantify the proportion of their interaction with external data, and classify the degree of smart contract dependency on external data. Researchers and developers can understand the proportions of external data dependency at different levels, enabling them to take appropriate measures for better practices.
    \item We identify and categorize five strategies for smart contracts to interact with external data (e.g., centralized/decentralized oracles, sidechain/cross-chain technology, and ZKPs), providing a systematic framework for understanding their implementation and domain-specific adoption patterns. This classification is supported by statistical analysis across key application domains such as DeFi, gaming, and supply chain management.
    \item We open-source our experimental data and codes on 
    \url{https://doi.org/10.6084/m9.figshare.25144934}
    
\end{itemize}

\section{BACKGROUND}
\subsection{Blockchain and Ethereum}
Blockchain \citep{004,farah2024survey,wu2024blockchain,moosavi2024blockchain}, as a decentralized and distributed ledger, securely stores data across a network of computers, safeguarding against unauthorized tampering and modification. The security of blockchain has facilitated its widespread adoption across finance, healthcare, and various other sectors. Ethereum, a blockchain-based platform, empowers developers to craft and implement smart contracts and DApps. Established by Vitalik Buterin and his team in 2015, Ethereum introduced the programmable blockchain, enabling the storage and execution of both transactions and executable code. Ethereum's flexible, Turing-complete programming environment positions it as developers' top choice for constructing decentralized solutions.

\subsection{Smart Contracts and Audit Reports}
Smart contracts \citep{005,liu2024overview,wu2024comprehensive} are Turing-complete programs executed on blockchain infrastructure. They are commonly implemented using languages such as Solidity, providing functionalities applicable across diverse real-world domains, including finance, cryptography, and healthcare \citep{006}. Their widespread deployment in financial applications renders them a frequent target for hacking endeavors. Owing to their immutable character post-deployment, security firms provide audit services to mitigate associated risks. Smart contract security audits \citep{007,xiao2024advanced,landsman2025auditing} encompass manual analyses conducted by specialized security teams to identify and assess potential vulnerabilities. Following their execution on simulated blockchains, audit reports are disseminated, encompassing details such as contract addresses, user permissions, specific vulnerabilities, and corrective recommendations \citep{008}. These reports are conventionally presented in English to accommodate a global readership.

\subsection{External Data Dependencies}
The external data dependency \citep{009} of smart contracts pertains to scenarios where a smart contract necessitates interaction with data sources external to the blockchain during execution. Typically, smart contracts necessitate the retrieval of data from external sources. This retrieval of external data is essential for smart contract execution, enabling them to make decisions predicated on real-time or dynamic external conditions. Contracts engage with off-chain data via Oracle services. Oracle services \citep{010,deng2024safeguarding} for smart contracts function as intermediaries, supplying external data to blockchain-based smart contracts and facilitating communication with real-world data. Furthermore, cross-chain communication \citep{011,falazi2024cross,cheng2024secure} is a prevalent method for smart contracts to interact with external data, frequently employing technologies such as cross-chain bridges, notarization mechanisms, relay chain protocols, and others. In this article, we present several real-world instances where smart contract interactions with external data have precipitated security concerns.

\section{METHODOLOGY}
\subsection{Overview}
\begin{figure*}[htbp]
    \centering
    \includegraphics[width=1\textwidth]{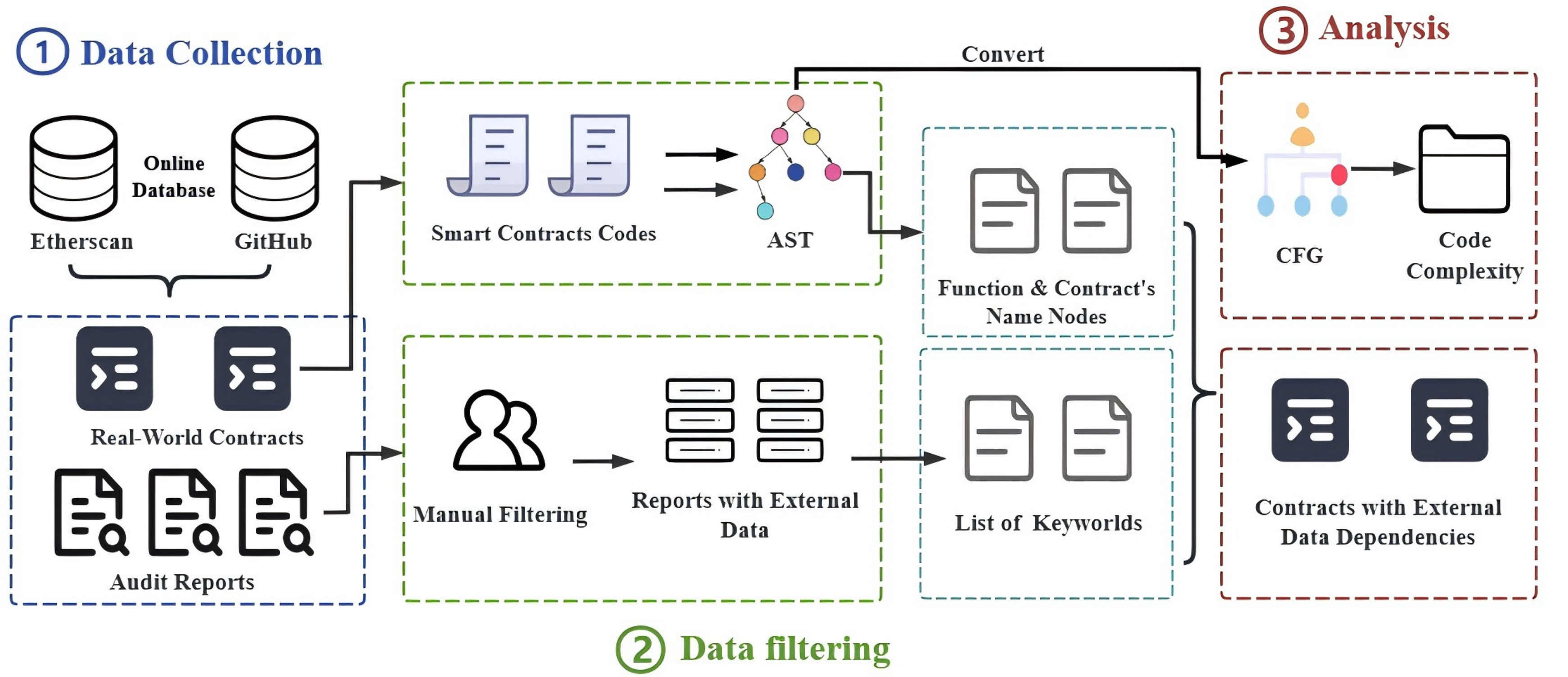}
    \caption{Investigative Process of Collecting and Analyzing Smart Contract Cases}
    \label{fig:1}
\end{figure*}

In this paper, we concentrate on the use of external data in real-world smart contracts. As depicted in \textcolor{blue}{Fig. \ref{fig:1}}. We have outlined a three-step process to calculate and analyze the use of external data in smart contracts. Firstly, we identified three distinct security teams offering auditing services on Etherscan. Thereafter, we conducted a manual review of their homepage information and downloaded over 3600 open-source audit reports in bulk, with some reports sourced from the security teams' websites and others from GitHub. Among these reports, a subset contained contract source codes, yielding a total of 240 contract source codes that featured interactions with external data. Following manual analysis, we compiled a keyword list encompassing contract and function names associated with external data dependencies. Employing web crawling techniques, we retrieved 10,500 real-world smart contracts from Etherscan. By parsing and comparing the contracts against the keyword list, we identified a total of 286 contracts relevant to external data interaction (with the statistics confined to successfully compiled Solidity projects). Below, we elaborate on the details of each step.

\subsection{Data collection}
To compile a dataset of real-world smart contracts, we employed web crawling technology. Constrained by the limitations of Etherscan's contract query interface, which permits access to only the most recent 500 contract addresses, we addressed this constraint by integrating a tool named SmartContractSpider into our project. This tool facilitated the collection of the 10,000 most recent contract addresses from Etherscan and the retrieval of the corresponding contract source code. The dataset of smart contracts compiled includes data through August 2023.\par
To gather security audit reports for these smart contracts, we identified three distinct security teams on Etherscan that provide contract auditing services. We located the official websites of these teams on their Etherscan profiles and found that two of the teams had published their audit reports as open-source on GitHub. Collectively, we obtained over 3600 audit reports from both official websites and GitHub repositories. These reports encompass data spanning from 2018 to August 2023.

\subsection{Data filtering}
To determine the number of contracts reliant on external data, we needed to exclude those utilizing Oracle services or involved in cross-chain communication. Initially, we conducted a manual review of 3600 security audit reports to identify those exhibiting external data access capabilities, yielding 249 reports that met the criteria. Among these 249 reports, we identified 240 open-source contract codes. Subsequently, we performed a manual analysis of these smart contracts to compile a keyword list titled "oracle\_services," which includes contract and function names potentially linked to external data access functionalities.\par

From the 249 audit reports that were filtered for containing external data access services, we performed a manual analysis. Subsequently, we calculated the frequency with which each smart contract project accessed external data across the reports. \textcolor{blue}{Table \ref{Table1}} presents the tally of contract and function names that include the relevant keywords, as compiled statistically.

\begin{table}
\centering
\begin{tabular}{l c c c} 
 \hline
 Categories & Method Name & Usage Count &  \\ 
 \hline
 Oracle Services & oracle	& 191 \\ 
 \hline
 crossChain & bridge & 39 \\
\hline
Oracle Services & chainlink & 13 \\
\hline
Oracle Services & external & 6 \\
\hline
crossChain & api & 22 \\
\hline
Oracle Services & dydx & 9 \\
\hline
crossChain & crosschain & 6 \\
\end{tabular}
\caption{Usage Count by Method Name and Category
}
\label{Table1}
\end{table}
\subsection{Analysis}
In the course of our research, we performed an extensive analysis of over 10,500 smart contracts utilizing ASTs. This process involves a detailed examination of the contracts' codebase to reveal its underlying structure and syntactic composition. Subsequently, we carefully extracted specific nodes from the ASTs that contained relevant information regarding the nomenclature of contracts and the functions they included and then stored these in a comprehensive repository referred to as 'total\_list'.\par
The subsequent phase of our investigation entailed a systematic comparison of the 'total\_list' with a curated collection of 'oracle\_services,' aimed at identifying potential correlations or commonalities. To ascertain the prevalence of specific keywords within the context of contract and function nomenclature, we conducted a rigorous comparative analysis,  Our empirical findings indicate that, of the 10,500 smart contracts analyzed, a subset of 286 contracts, representing approximately 2.86\% of the sample, were found to enable interactions with external data sources. \par
Concurrently with our ASTs analysis, we acquired and analyzed 249 security audit reports pertaining to smart contracts. These reports offer valuable insights into the frequency of external data access across diverse projects. Our statistical analysis yields a spectrum of external data access frequencies, with the maximum number of accesses recorded at an astonishing 11,724 instances. The average frequency of external data access among the audited projects was calculated to be 117 occurrences per project. To elucidate the patterns of external data access within smart contracts, we utilized a visualization technique to depict the distribution of access frequencies graphically. Based on the observed distribution, we applied a decision tree classification method \cite{marudi2024decision,mienye2024survey,alnuaimi2024overview} to establish two critical thresholds, facilitating the categorization of access frequencies into three distinct levels: low, medium, and high. This stratification approach not only offers a clear overview of project distribution but also enhances the understanding of the dependency on and variability of external data within the smart contract ecosystem.\par

\begin{figure*}[t!]
\centering
\includegraphics[width=0.68\textwidth]{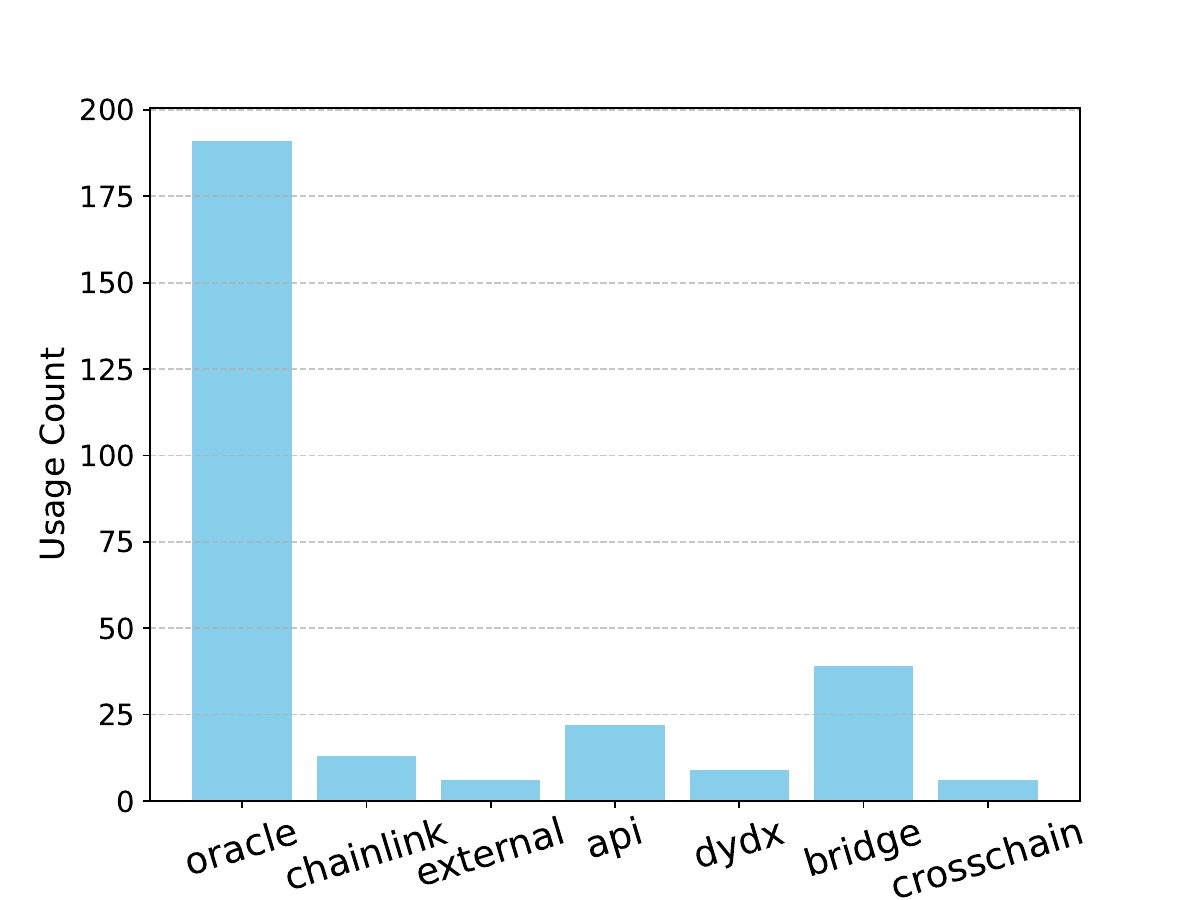}
\caption{The Count of Function Name or Contract Name}
\label{fig:2}
\end{figure*}

\begin{figure*}[t!]
\centering
\includegraphics[width=0.68\textwidth]{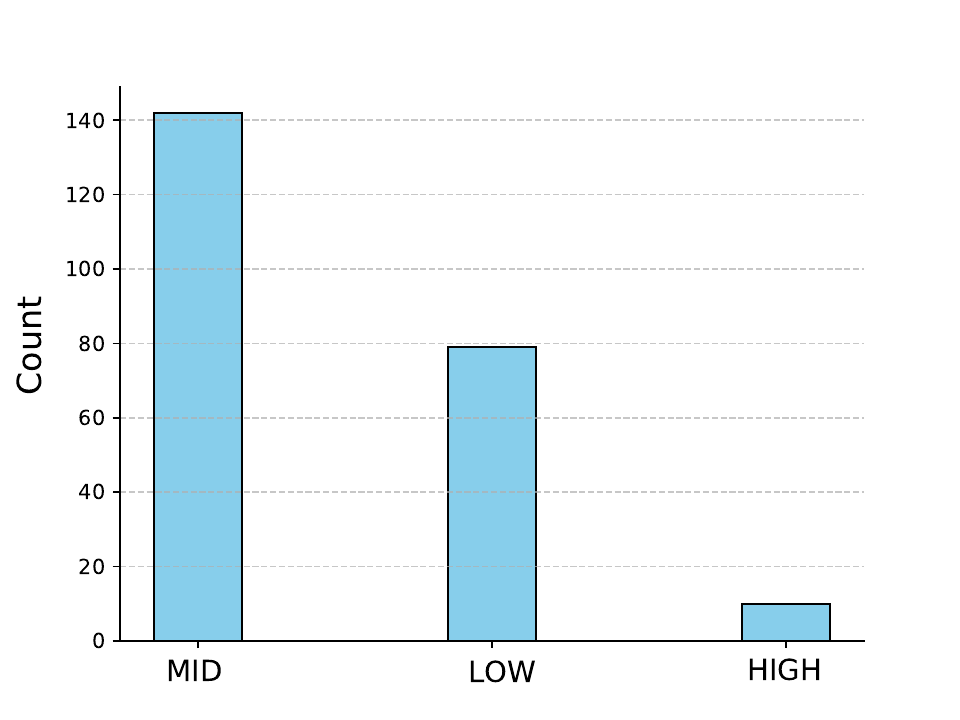}
\caption{The Frequency of Access Frequency Levels to External Data}
\label{fig:3}
\end{figure*}

Utilizing this methodology, we systematically document and present the number of projects within each frequency category and the count of function names or contract names as visually represented in \textcolor{blue}{Fig.\ref{fig:2}} and \textcolor{blue}{Fig.\ref{fig:3}} The distribution of projects across these categories provides a comprehensive understanding of the prevalence and variability of external data reliance within the smart contract ecosystem.\par

\begin{algorithm}[htp!]
\caption{Calculate Cyclomatic Complexity of Solidity Smart Contracts}
\label{Algorithm:1}
\begin{algorithmic}[1]

\STATE \textbf{Function} build\_cfg\_from\_ast(ast)
\STATE \hspace{1em} \textbf{Create} directed graph cfg
\STATE \hspace{1em} \textbf{Define} recursive function parse\_node(node, parent)
\STATE \hspace{2em} \textbf{If} node is a dictionary \textbf{then}
\STATE \hspace{3em} Get node type node\_type and generate node\_id
\STATE \hspace{3em} \textbf{Add} node and edge to cfg
\STATE \hspace{3em} Recursively parse conditions and bodies of control statements
\STATE \hspace{3em} Recursively parse block and function bodies, and other child nodes
\STATE \hspace{2em} \textbf{If} node is a list \textbf{then} recursively parse each item
\STATE \hspace{1em} Parse AST and return cfg

\STATE \textbf{Function} calculate\_cyclomatic\_complexity(cfg)
\STATE \hspace{1em} Get num\_edges, num\_nodes, num\_connected\_components
\STATE \hspace{1em} \textbf{Calculate} cyclomatic complexity as num\_edges - num\_nodes + 2 * num\_connected\_components
\STATE \hspace{1em} Return cyclomatic complexity

\end{algorithmic}
\end{algorithm}

\begin{figure*}[t!]
\centering
\includegraphics[width=0.70\textwidth]{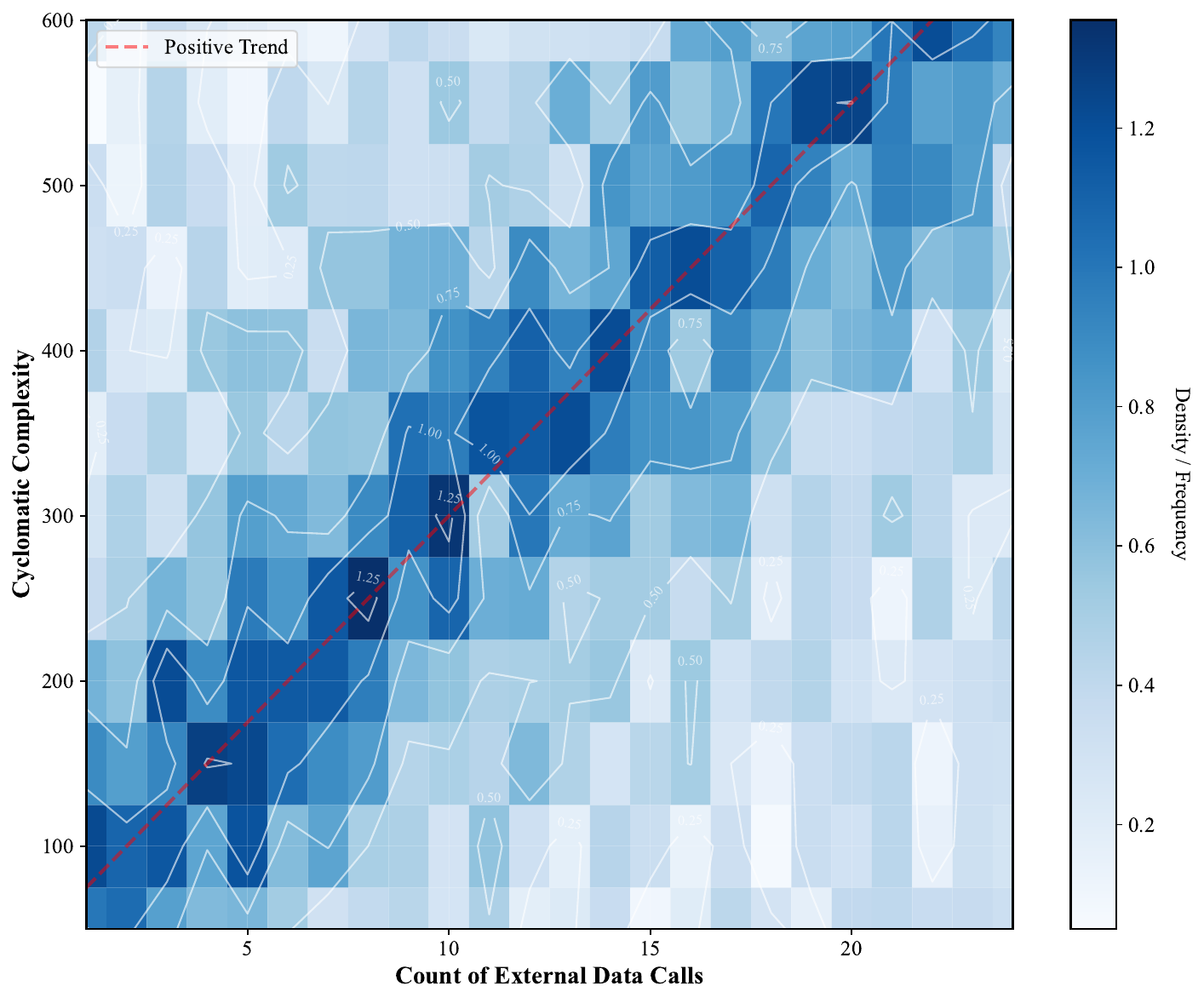}
\caption{The Count of External Data Calls and Code Complexity}
\label{fig:4}
\end{figure*}

Subsequently, we compiled and analyzed 74 smart contracts from a total of 249 audit reports, transforming them into ASTs and subsequently into CFG. Applying the cyclomatic complexity formula: $V(G) = E - N + 2$ (where $E$  represents the edges and $N$ represents the nodes in the CFG), the process is detailed in the accompanying \textcolor{blue}{Algorithm \ref{Algorithm:1}}. This methodology addresses the challenge posed by the lack of existing libraries for computing the cyclomatic complexity of Solidity contracts. We observed a significant positive correlation (with a correlation coefficient $r$ and a $p - value < 0.5$) between the number of external data dependencies and the code complexity of smart contracts, as depicted in \textcolor{blue}{Fig.\ref{fig:4}}

Within the range of low dependency counts, the increase in code complexity is relatively gradual; however, with an increasing number of dependencies, the code complexity exhibits a trend of accelerating growth. Several factors may contribute to the observed positive correlation\citep{012}. Firstly, an increased dependence on external data may require incorporating more complex logical judgments and state management mechanisms within the contract. Secondly, to process and verify external data, the contract may need to include a more significant number of conditional statements and loop structures, thereby increasing the cyclomatic complexity of the code.

\section{Smart Contract Domain Distribution and Data Interaction Strategies}
\subsection{Distribution of Smart Contracts Across Domains}
We categorized the 9,356 collected contracts manually, revealing that they are predominantly utilized in decentralized finance, gaming, and supply chain management. Additionally, contracts are applied, albeit to a lesser extent, in IoT, insurance, identity authentication, education, and other sectors. Next, we provide specific information on each of the main application areas of smart contracts.

\subsubsection{\textbf{DeFi}}
\ 
\newline
DeFi\citep{013} is among the most prominent applications of smart contract technology. Decentralized oracles are predominant in this sector, supplying dependable external data for decentralized exchanges, lending platforms, and other financial instruments. Moreover, sidechain and cross-chain technologies are crucial for lowering transaction costs and enabling asset interoperability among different blockchain networks. With escalating privacy concerns, ZKPs technology is increasingly adopted in the DeFi sector, particularly for scenarios that necessitate safeguarding transaction details and account balances.

\subsubsection{\textbf{Supply Chain Management}}
\
\newline
In supply chain management\citep{014}, smart contracts are mainly used to optimize logistics and inventory systems. High transaction and data processing demands have led to the widespread adoption of sidechain technology, which effectively reduces the load on the main blockchain and enhances system efficiency. Furthermore, smart contracts in this domain heavily depend on Oracle technology to obtain external data like market prices and shipping statuses, ensuring accuracy and real-time responsiveness. Though the use of  ZKPs is still limited, their capacity to bolster data privacy is gaining traction, especially in scenarios with sensitive commercial data.

\subsubsection{\textbf{Gaming}}
\
\newline
The gaming industry has a significant need for external data interaction\citep{015}, particularly for event triggers, random number generation, and verifying player behavior. Centralized oracles are essential in this domain, providing an efficient and simple method for accessing external data. Sidechain technology is often employed in blockchain-based gaming ecosystems to alleviate the main blockchain's load and reduce transaction costs. Despite their limited prevalence in the gaming sector, cross-chain technology and ZKPs show promise in certain scenarios, including facilitating interoperability across various blockchain networks and safeguarding sensitive gaming data.

\begin{figure*}[htbp]
    \centering
    \includegraphics[width=0.95\textwidth]{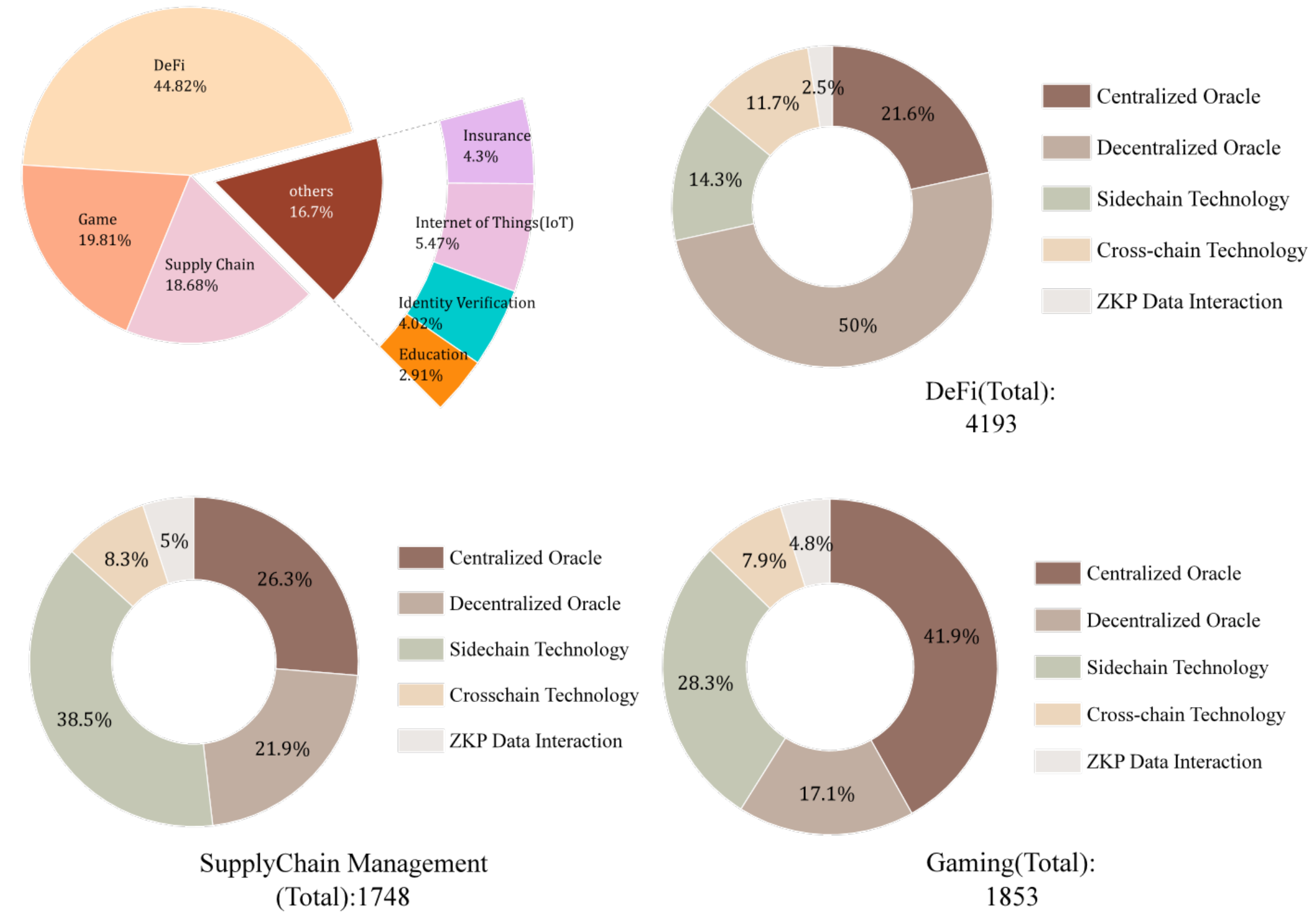}
    \caption{Smart Contract Distribution \& External Data Interactions}
    \label{fig:5}
\end{figure*}

\subsection{Strategies for Smart Contract Interaction with External Data}
Given the low accuracy of current automated tools in filtering external data interaction functionalities, we refined our automated screening strategy and performed additional manual analysis on smart contracts involving external data interaction. Through extracting and analyzing the functional methods and libraries within the source code of each smart contract, we discerned five distinct strategies for interacting with external data. Subsequently, we statistically analyzed the distribution of these five strategies among smart contracts in various domains. The statistical application of smart contracts in various fields, as well as the statistics of five types of interactions with external data in each field, are shown in \textcolor{blue}{Fig. \ref{fig:5}} Additionally, we detailed the processes for each strategy and extracted the key code features of each approach. This serves as a reference for future developers and researchers to directly utilize in similar studies.

\subsubsection{\textbf{Centralized Oracle}}
\
\newline
Centralized oracles\citep{016}, as singular entities, supply external data to blockchain networks and act as the authoritative sources for information like market prices and real-world events. Despite their efficiency and simplicity, centralized oracles present a single point of failure and potential trust issues because their data feeds can be manipulated or experience outages. Relying on a single data source can compromise the decentralized nature of blockchain systems. \par
The centralized oracle serves as an intermediary within the blockchain system, tasked with integrating external data into on-chain smart contracts. The typical implementation process encompasses the following stages\citep{017}: Initially, the oracle retrieves necessary information from external sources such as APIs, databases, or sensors, ensuring the data's completeness and timeliness during collection. Next, the oracle validates the collected data for accuracy and completeness, serving as a reliable data source for executing smart contracts. Following this, the verified data is transmitted to the blockchain network via the oracle server, thereby finalizing the on-chain data transmission and dissemination. Eventually, the smart contract receives the data provided by the prediction machine and executes the corresponding operation according to the prediction logic, thus realizing the efficient interaction between external information and the blockchain system. \par
The sample code is illustrated in \textcolor{blue}{Listing \ref{lst:Listing1}}. Within this contract, the oracle address acts as the centralized predicate machine, and it alone can execute the 'updateData' function to modify the data on-chain. However, centralized oracles are prone to a single point of failure and trust issues, potentially compromising the decentralized nature and security of blockchain systems\citep{018}. Thus, despite the simplicity of implementing centralized oracles, their associated risks must be balanced against system requirements in practical applications.
\vspace{7pt}
\begin{lstlisting}[language=Solidity,caption={The Sample Code of Centralized Oracle},
label={lst:Listing1}]
contract CentralizedOracle {
    address public oracle;
    uint256 public data;
    modifier onlyOracle() {
        require(msg.sender == oracle, "Not authorized");
    }
    constructor(address _oracle) {
        oracle = _oracle;
    }
    function updateData(uint256 _data) public onlyOracle {
        data = _data;
    }
}
\end{lstlisting}

\subsubsection{\textbf{Decentralized Oracle}}
\ 
\newline
Decentralized oracles\citep{019} reduce the risks of centralization by consolidating data from various sources, which increases reliability and trustworthiness. They use consensus mechanisms to authenticate information prior to its transmission to smart contracts, minimizing the chance of data manipulation. This approach is in line with blockchain's foundational principles, ensuring that no single entity exerts excessive control over the data. \par
During implementation, a smart contract initially requests specific data, received by several independent oracle nodes. In response to the request, each node gathers the necessary information from its designated external data sources and then submits this data to the blockchain. Subsequently, the data submitted by all the nodes are verified and aggregated through a predetermined consensus mechanism (e.g., majority voting, weighted average, etc.) to ensure the accuracy and integrity of the data. Ultimately, the consolidated data results are transmitted to the smart contract, triggering the execution of its logic and facilitating the secure interaction between the blockchain system and external data\citep{020}.\par
The sample code is illustrated in \textcolor{blue}{Listing \ref{lst:Listing2}.} Within this contract, the 'AggregatorV3Interface' interface facilitates interaction with decentralized oracles, including Chainlink's data aggregator, to retrieve the most recent price data. Decentralized oracles mitigate trust risks and enhance data reliability and accuracy by incorporating multiple independent data sources and nodes\citep{021}. As noted in the Chainlink whitepaper, “Existing oracles rely on a centralized service model, which can lead to a single point of failure in smart contracts, yet they are considered more secure than traditional centralized digital contracts.”

\vspace{7pt}
\begin{lstlisting}[language=Solidity,caption={The Sample Code of  Decentralized Oracle},
label={lst:Listing2}]
interface AggregatorV3Interface {
    function latestRoundData()
    external
    view
    returns(
        uint80 roundID,
        int256 answer,
        uint256startedAt,
        uint256 updatedAt,
        uint80 answeredInRound
    );  
}
contract DecentralizedOracleConsumer {
    AggregatorV3Interface internal priceFeed;
    constructor(address _aggregator) {
        priceFeed = AggregatorV3Interface(_aggregator);
    }
    function getLatestPrice() public view returns (int256) {
        (int256 price,) = priceFeed.latestRoundData();
        return price;
    }
}
\end{lstlisting}

\subsubsection{\textbf{Sidechain Technology}}
\ 
\newline
Sidechains\citep{022} are separate blockchains that function alongside a mainchain, facilitating the transfer of assets and data between them. They promote scalability and flexibility by transferring transactions away from the mainchain, thereby reducing network congestion and improving performance. Sidechains can be tailored to specific use cases, offering a sandbox for testing new features without endangering the security of the main network. The typical implementation process encompasses the following steps\citep{023}:
\begin{itemize}
    \item \textbf{Locking Assets}: Users lock a specific quantity of assets on the main chain, typically via a smart contract, which generates a proof that the assets are frozen and inaccessible on the main chain.
    \item \textbf{Proof Verification}: The sidechain confirms the legitimacy of the locking operation and validates that the appropriate number of assets are secured on the main chain.
    \item \textbf{Minting Assets}: Upon successful verification, the sidechain mints tokens or assets of equivalent value to the locked assets, making them available for use on the sidechain.
    \item \textbf{Sidechain Operations}: Users conduct a range of operations on the sidechain, including transactions and smart contract executions, to leverage the unique features and benefits it offers.
    \item \textbf{Redeeming Assets}: To transfer assets from the sidechain to the main chain, users must burn the equivalent value of tokens on the sidechain and produce the corresponding proof.
    \item \textbf{Unlocking Assets}: Once the main chain confirms the proof's validity, it releases the previously locked assets, reinstating their availability.
\end{itemize}
The sample code is illustrated in \textcolor{blue}{Listing \ref{lst:Listing3}.}
\vspace{8pt}
\begin{lstlisting}[language=Solidity,caption={The Sample Code of  Sidechain Technology},
label={lst:Listing3}]
contract MainchainLock {
    mapping(address => uint256) public lockedAssets;
    function lockAssets(uint256 amount) public {
        require(amount > 0, "Amount must be greater than zero");
        //Assuming the user has approved the contract to transfer their tokens
        //Token Transfer Function
        lockedAssets[msg.sender] += amount;
        //Generate a proof of lock (in a real implementation, events or other mechanisms may be involved)
    }
}
\end{lstlisting}

On the sidechain, the necessary verification and minting logic must be implemented to ensure the secure transfer of assets. Notably, sidechain implementation must address the complexities of cross-chain communication, security, and consensus mechanisms. For instance, the Zendoo protocol offers a verifiable cross-chain transfer protocol utilizing zk-SNARKs, enabling the main chain to engage with various sidechains without requiring knowledge of their internal structures\citep{024}. The protocol generates succinct proofs of sidechain state progression through the recursive combination of zk-SNARKs, allowing the main chain to efficiently validate sidechain transactions.

\subsubsection{\textbf{Cross-chain Technology}}
\
\newline
Cross-chain technology\citep{025} enables interoperability among various blockchain networks, allowing for the frictionless transfer of assets and data across different platforms. Such interoperability is essential for fostering a unified blockchain ecosystem, enabling users to harness the unique advantages of different networks. Methods like atomic swaps, relays, and decentralized oracles are used for cross-chain communication, each presenting distinct trade-offs in terms of security, speed, and complexity.\par
Cross-chain technology is designed to facilitate interoperability among diverse blockchain networks, enabling the frictionless transfer and interaction of assets and data across various chains.  The typical implementation process encompasses the following steps\citep{026}:
\begin{itemize}
    \item \textbf{Cross-chain communication protocols}: Protocols are standardized to facilitate secure and reliable information exchange between disparate blockchain networks \cite{miyaji2024efficient} . They stipulate message formats, transmission mechanisms, and authentication procedures.
    \item \textbf{Atomicity of Cross-Chain Transactions}: Ensure that cross-chain transactions are atomic, meaning they either execute successfully across all involved chains or are fully rolled back to avoid partial completions that could result in inconsistencies \cite{deng2025enhancing}. This is commonly achieved using atomic swaps or complementary protocols.
    \item \textbf{Cross-chain smart contracts}: Develop smart contracts capable of executing across various blockchains to manage cross-chain operations. These contracts must be able to initiate cross-chain calls and synchronize states
    \item \textbf{Relay and Verification Mechanism}: Implement relay nodes or verifiers tasked with monitoring various blockchains' statuses and facilitating the transfer and verification of transaction data between chains, ensuring the security and precision of cross-chain operations \cite{xu2024relay} .
    \item \textbf{Bridging}: Establish a mechanism for cross-chain bridging that enables the transfer of assets between distinct blockchains \cite{zhang2024security} . This process generally involves locking assets on the originating chain and issuing an equivalent token on the destination chain.
\end{itemize}

The sample code is illustrated in \textcolor{blue}{Listing \ref{lst:Listing4}}.
\vspace{8pt}
\begin{lstlisting}[language=Solidity,caption={The Sample code of Cross-chain Technology},
label={lst:Listing4}]
contract CrossChainBridge {
    mapping(address => uint256) public lockedBalances;
    event Locked(address indexed user, uint256 amount, string targetChain);
    function lockTokens(uint256 amount, string memory targetChain) public {
        require(amount > 0, "Amount must be greater than zero");
        // Assuming the use of ERC20 standard tokens
        // Token Transfer Function
        lockedBalances[msg.sender] += amount;
        emit Locked(msg.sender, amount, targetChain);
        // In a practical implementation, it may be necessary to interact with the bridging contract of the target chain
    }
}
\end{lstlisting}

\subsubsection{\textbf{ZKPs (Zero-Knowledge Proof) Data Interaction}}
\ 
\newline
ZKPs \citep{027,lavin2024survey,xing2025zero,gupta2025zero} are cryptographic techniques that enable a party to verify the truth of a statement without disclosing extra information. In blockchain applications, ZKPs bolster privacy and security by allowing verification of data without exposing the underlying data. This is especially advantageous in confidential scenarios like identity verification and private transactions, ensuring sensitive data stays safeguarded during necessary validation processes. \par
In implementing a ZKPs protocol, the prover and verifier initially agree on the protocol to be used and establish the required public parameters, ensuring the protocol's framework and fairness. Within the protocol's framework, the prover must define the statements to be proven and translate them into mathematical forms compatible with ZKPs for further processing. Subsequently, the prover creates a proof of a specific mathematical model or rule, a process that often requires intricate calculations to validate the statement's truth without revealing sensitive data. The prover subsequently submits the proof to the verifier for validation. Finally, the verifier assesses the proofs using the protocol's specified verification methods to ascertain the statement's validity and ensure compliance with the ZKPs protocol's requirements\citep{028}. \par
The sample code is illustrated in \textcolor{blue}{Listing \ref{lst:Listing5}}.
\vspace{8pt}
\begin{lstlisting}[language=Solidity,caption={The Sample Code of  ZKPs},
label={lst:Listing5}]
contract ZKProofVerifier is Verifier {
    function verifyProof(
        bytes memory proof,
        uint256[1] memory input
        ) public view returns (bool) {
            return verify(proof, input);
        }
}
\end{lstlisting}

\subsubsection{\textbf{Complementary}}
\begin{figure*}[htbp]
    \centering
    \includegraphics[width=0.90\textwidth]{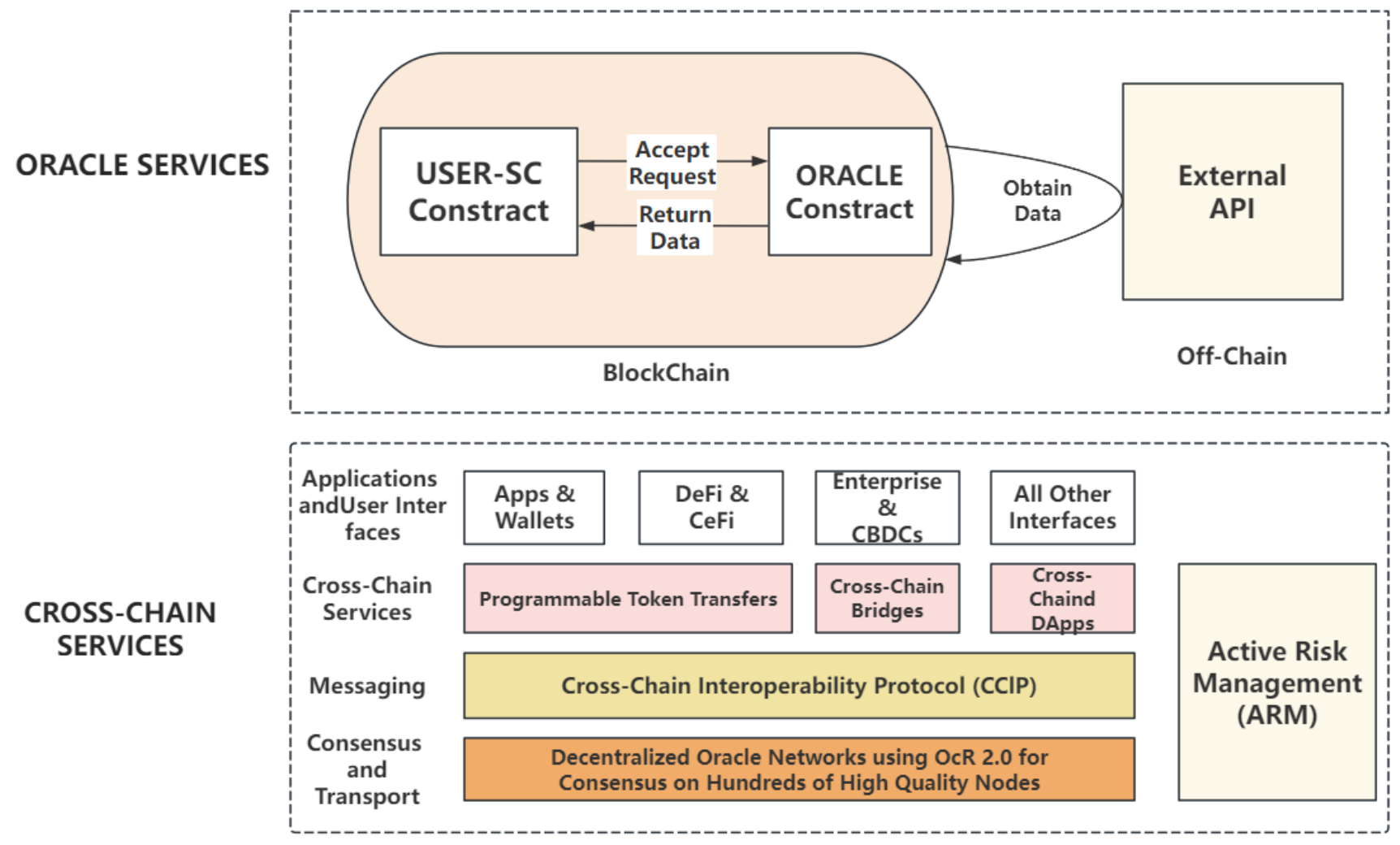}
    \caption{Oracle Services and Cross-Chain Services}
    \label{fig:6}
\end{figure*}

\ 
\newline
We conduct an extensive review of the current literature and the collected contract audit reports, synthesizing the potential security issues arising from the interaction of smart contracts with external data into two distinct yet interrelated areas: issues related to the credibility of the data used, and security challenges associated with the use of Oracle services, as described in \textcolor{blue}{Fig.\ref{fig:6}}.

Data trustworthiness issues \cite{nguyen2025blockchain} encompass both the inherent insecurity of the data itself and the possibility of data tampering, both of which can lead to the contamination of the data source. The execution of smart contracts relies on external data sources. However, an insecure data source may pose safety hazards to the smart contracts that depend on such data \citep{029}. \par
If the Oracle network fetches data from only one trading platform API, it cannot provide effective protection during that platform's downtime, flash crashes, or price manipulation. This situation leads to the risk of smart contracts executing based on incorrect price data, potentially resulting in the loss of user funds. Data auditing is a primary method to ensure the security of data sources, but auditing external data by a third-party auditor (TPA) \cite{zhao2024smart,miao2024blockchain} authorized by public auditing still presents security concerns, as TPA is often considered semi-honest. To address this issue, Kuan et al. proposed a decentralized auditing scheme (Dredas) by replacing the TPA with a designed smart contract. In this scheme, anyone can obtain audit results from Ethereum without worrying about a semi-honest TPA. Kuan's article provides a solution to the external data trustworthiness issue by shifting data auditing from off-chain to on-chain. By designing smart contracts with data auditing capabilities, external data is audited on-chain before being utilized. \par
Oracles serve as a bridge for blockchains to access external data and multiple DeFi projects within the Ethereum network have encountered oracle issues, with some being attacked due to their reliance on a single oracle. Kevin et al. \citep{030} have identified a manipulation process where attackers exploit unsecured flash loans to manipulate market prices provided by oracles, thereby illicitly profiting within smart contracts.

\section{RELATED WORK}
With the development of blockchain technology, the security of smart contracts has increasingly become a focal point. Several insightful studies have been proposed to investigate security issues within blockchain and smart contracts. Rafique et al. \citep{031} provided a detailed introduction to the concept of Oracle services and elucidated the process through which smart contracts utilize the Oracle mechanism to obtain off-chain data. Dong et al. \citep{032} point out that Oracle services always risk providing damaged, malicious, or inaccurate data. They introduce the Distributed Autonomous Oracle Network (DAON), its consensus protocol, and non-interactive schemes for reputation maintenance and payments. Vinayasree et al. \citep{033} propose solving query performance issues and limited query semantics by linking to other databases. They achieve version control functionality by independently designing version control semantics. Luo et al. \citep{034}. researched and describe widely used blockchain Oracle services, providing detailed insights into its potential roles, technical architecture, and design patterns.
\section{CONSLUSION}
In this paper, we comprehensively analyzed how smart contracts depend on external data, providing empirical insights into its significance and impact. Our study involved gathering and analyzing a large dataset of real-world smart contracts, uncovering the widespread and critical nature of their interaction with external data. This analysis has shed light on potential issues arising from this dependency, providing quantified data on its extent and categorizing the degree of dependence to guide developers' decision-making. \par
Our research has made several important contributions to the field. Firstly, we have conducted the first empirical study on the dependence of smart contracts on external data, addressing a critical gap in the existing literature. This study helps to reduce security issues caused by developers' interactions with external data in smart contracts. Secondly, we have analyzed real-world smart contracts, quantifying the proportion of their interaction with external data and classifying the degree of dependence. This information is valuable for researchers and developers to understand the proportions of external data dependency at different levels and take appropriate measures for better practices. Finally, we have open-sourced our experimental data and codes to facilitate further research in this area. \par
By understanding the common strategies and vulnerabilities associated with external data interaction, developers can design more secure and reliable smart contracts. Auditors, on the other hand, can use this knowledge to conduct more thorough and targeted security assessments. Looking forward, several promising directions for future research have emerged. Further investigation into the credibility of external data sources and the development of more robust security auditing techniques are essential for ensuring the continued growth and stability of blockchain applications. Additionally, exploring innovative methods to mitigate the risks associated with external data dependencies, such as advanced encryption and decentralized data verification mechanisms, could significantly enhance the security and reliability of smart contracts in real-world scenarios.

\end{document}